\documentclass[twocolumn,prb,showpacs,preprintnumbers,amsmath,amssymb]{revtex4}
\usepackage{graphicx}
\usepackage{mathptmx}
\usepackage{verbatim}

\begin{document}
\title{Gapped broken symmetry states in ABC trilayer graphene}
\author{Jeil Jung} \email{jeil@physics.utexas.edu}
\affiliation{Department of Physics, University of Texas at Austin, USA}
\author{Allan H. MacDonald} 
\affiliation{Department of Physics, University of Texas at Austin, USA}
\date{\today{}}
\begin{abstract}
We use a self-consistent Hartree-Fock approximation
with realistic Coulomb interactions for $\pi$-band electrons 
to explore the possibility of broken symmetry states in 
weakly disordered ABC stacked trilayer graphene.  
The competition between gapped and gapless broken symmetry states,
and normal states is studied by comparing total energies.
We find that gapped states are favored and that, unlike the 
bilayer case, gapless nematic broken symmetry states are not metastable.
Among the gapped states the layer antiferromagnetic state is favored over anomalous
and spin Hall states.
\end{abstract}
\maketitle

\noindent
\section{Introduction}
The electronic structure of few layer graphene\cite{multi,hongki1} systems consists of pairs of 
bands which cross, or narrowly avoid crossing, near the Fermi level.  Because the number of 
band pairs depends in an interesting way on the stacking arrangement, 
and because both semi-metallic and semi-conducting behaviors occur, 
this family of two-dimensional materials provides an attractive playground for the study of electron interaction 
effects in systems\cite{abrikosov,dirachf} with approximate Fermi points.
For example, interactions lead to a marginal Fermi liquid behavior
in neutral single layer graphene,\cite{guinea,Manchester_Velocity} and to 
broken-symmetry ordered states in bilayers.
\cite{hongki,bilayers_gap,bilayers_nem,bilayers_bsymsurvey,jeilbilayer,fan,jairo,geim,yacoby}

The recent surge of interest in ABC stacked trilayer graphene 
\cite{trilayers1, trilayers2, trilayers3,abctri1,abctri2,abctri3,abctri4,abctri5,abctri6,abctri7,abctri8,abctri9}
motivates theoretical studies of the electron interaction induced instabilities that are  
expected when these structures are weakly disordered.  
Instabilities are favored in ABC trilayers by extremely flat crossing\cite{hongki1} of a single-pair 
of bands at the neutral system Fermi level, and by exchange energy frustration associated with 
mometum-space textures in the valence band wavefunctions.\cite{nobelsymposium}
In the trilayer case the competition between competing broken symmetry states is 
massaged by weak remote neighbor hopping processes that 
reshape the bands at energies within $\sim 20$ meV   
of the crossing point. \cite{trilayerbands} 
This energy scale should be compared to the 
$\sim 1$ meV scale of analogous processes in bilayer graphene \cite{mccann}.
The remote hopping processes are therefore more likely to 
play a prominent role in determining how the system responds to
electron-electron interactions
in the trilayer case.

The broken symmetry states that have been discussed
in the bilayer graphene literature can broadly be classified either as
gapped phases with broken layer inversion symmetry, \cite{hongki,jeilbilayer,fan,bilayers_gap} 
or as gapless nematic states that lower rotational symmetry. \cite{bilayers_nem}
Although the two types of states in principle should have clear experimental signatures,
it has not yet been possible\cite{jairo,geim,yacoby} to achieve a universally accepted consensus on the character of
the ground state because of the complicating role of residual disorder.
Studies of ABC stacked trilayer graphene could prove to be more unambiguous 
because its bands are flatter and interaction effects correspondingly stronger,
while disorder strengths should be comparable.

In this paper we present a study of the competition between gapful and gapless states in ABC stacked 
trilayer graphene, including the effects of weak remote-hopping processes which 
can dominate band dispersion very close to the band-crossing (Dirac) point.
Our study is based on a six-band $\pi$-orbital tight-binding model, 
combined with long-range Coulomb interactions treated using a Hartree-Fock 
mean-field-theory.  The quasiparticle band-structures of both gapped and gapless states  
are reshaped when interactions are included.  
We find that gapped phases are favored over a wide range
of the hopping-parameter model space.  In mean-field theory
the energy difference between gapped and gapless states is  
typically smaller than $\sim 10^{-7}$ eV per carbon atom.
The small condensation energy reflects the fact that only single-electron states close to the 
band crossing points participate in ordering.  The strength of the direct hopping process between
low-energy sites on the outer layers of the trilayer, $\gamma_2$, plays an especially
important role in determining the character of the ground state.

\begin{figure*}[t]
\begin{center}
\includegraphics[height=4cm,angle=0]{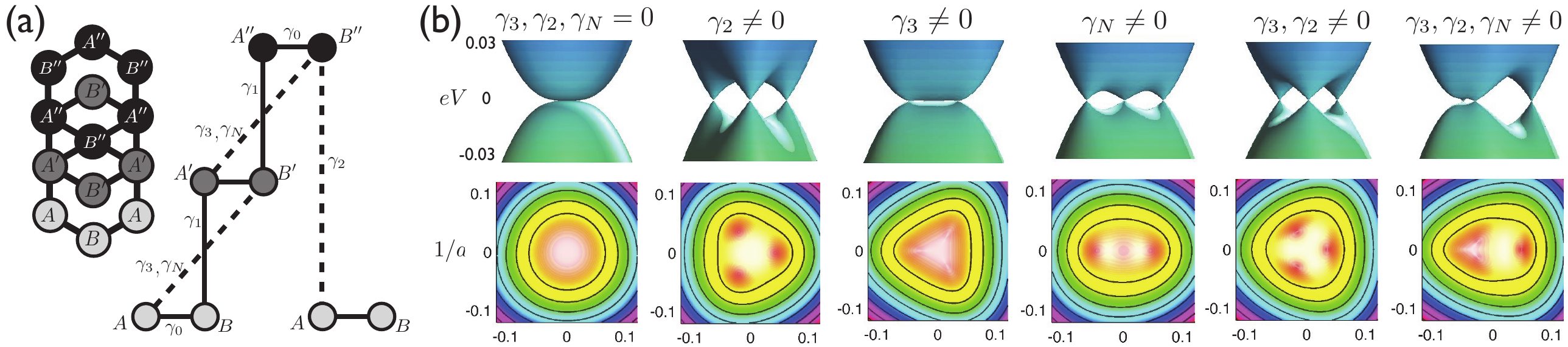}
\caption{
Trilayer graphene unit cell and $\pi$-band structure.
(a) Schematic representation of the hopping processes included in our model.
(b) Band structure near the Dirac point $K = (4\pi/3a,0)$ for different values of the 
remote hopping parameters.
The upper row shows 2D band structures seen from a view rotated by 30$^{\circ}$ with 
respect to vertical while the lower row shows the same information expressed in terms of 
contour plots.  When non-zero, the hopping parameters have the values  
$\gamma_2 = 0.01$ eV, $\gamma_3 = 0.3$ eV.  We have set 
$\gamma_4 = \gamma_5 = 0$ throughout our calculations.
Wave vectors $k$ are in units of $a^{-1}$.
The $\gamma_2$ term splits the Brillouin-zone corner cubic 
band crossing into three Dirac cones (linear band crossings) located at the vertices of an equilateral triangle.
The trigonal $\gamma_3$ term acting on its own results in four 
band-crossing points, including one at the Brillouin-zone corner.
When both terms are present simultaneously the gapless point at $K$ disappears.
When the signs of both terms are the same they tend to produce opposing triangular distortions, 
whereas if they have opposite signs their triangular distortions are reinforcing.
The orientation of the triangular distortion for each of the above parameters is sign dependent.
In realistic band structures the $\gamma_2$ term dominates over $\gamma_3$ resulting in three 
Fermi points with linear dispersion.
The nematic term $\gamma_N$ captures the influence of a layer relative-sliding 
sliding strain and breaks the triangular 
rotational symmetry of the bands. We used the value $\gamma_N^0 = 0.02$ eV in the above illustration.
}
\label{fig1}
\end{center}
\end{figure*}

\section{Model Hamiltonian}
We describe ABC trilayer graphene using a 
lattice model Hamiltonian with one atomic $2 p_z$ orbital per carbon site.
We label the six sublattice sites illustrated 
in Fig. 1(a)  {\em A}, {\em B}, $A^{\prime}$, $B^{\prime}$, $A^{\prime \prime}$, $B^{\prime \prime}$;  the 
$A$ and $B^{\prime \prime}$ sites avoid near-neighbor inter-layer coupling and for this reason are 
{\em low-energy sites} which are dominantly occupied by electrons close to the band crossing points.   
With this convention, the six band tight-binding model Hamiltonian of ABC trilayer graphene is:
\begin{equation}
\label{hamil}
{H}_0=  -
\begin{pmatrix}
     0                    &    \gamma_0 f         &   0 
      &   \gamma_3 f^* + \gamma_N  &  0    &   \gamma_2   \\
 \gamma_0 f^*    &      0                        &   \gamma_1      &   
 &  0    &   0 \\
 0 &       \gamma_1        &                     0   &   \gamma_0 f      &  0 
 &   \gamma_3 f^* \\
 \gamma_3 f   + \gamma_N^*    &   0 
 & \gamma_0 f^*   &    0    &      \gamma_1    &   0   \\
 0                        &   0                           &  0  
 &   \gamma_1     &     0    &   \gamma_0 f   \\
 \gamma_2                        &   0                           & \gamma_3 f                       &    0    &     \gamma_0 f^*    &   0   \\
\end{pmatrix}
\end{equation}
where
\begin{eqnarray}
f\left( {\bf k} \right) &=&    e^{ i k_y a / \sqrt{3} } \left( 1 + 2 e^{-i 3 k_{y} a / 2\sqrt{3}}  
                                       \cos \left(  \frac{k_x a}{2} \right)    \right)  
\end{eqnarray}
with $a = 2.46 \AA$ using the same triangular lattice vector convention as in Ref. [\onlinecite{dirachf,jeilbilayer}].
The global minus sign in front of the Hamiltonian means that $\pi$-bonding bands 
have lower energy than anti-bonding bands when the $\gamma$ parameters are positive.  
In most of our calculations we have used 
graphite hopping parameter values which are similar to those in Ref. [\onlinecite{partoens}] :
$\gamma_0 = 3.12$ eV, $\gamma_1 = 0.377$ eV,  $\gamma_2 = 0.01$ eV,  $\gamma_3 = 0.3$ eV.
We specifically address the importance of the signs of the remote $\gamma_2$ and 
$\gamma_3$ hopping parameters.  
The near-neighbor intralayer and interlayer hopping processes $\gamma_0$ and $\gamma_1$ 
are responsible for broad features of the band structure, while the 
$\gamma_2$ and $\gamma_3$ parameters have their 
main impact close to the band-crossing points.  
This model qualitatively reproduces the {\em ab initio} band structure in Ref. [\onlinecite{latil}],
in particular capturing the orientation of the triangle formed by the three 
band-crossing points close to the Brillouin-zone corner.
We have ignored the ABC trilayer 
 $\gamma_4$ and $\gamma_5$ processes that break particle-hole symmetry, 
and other small onsite terms that are often introduced in models of graphite, because 
they do not visibly alter the low energy features of the bands in ABC trilayer graphene.

Using a model similar to that used previously for bilayer graphene,\cite{youngwoo,kruczynski}. 
we have also examined the influence of a term in the Hamiltonian that is intended to capture the 
influence on low-energy states of an interlayer relative-translation strain.  
We write 
$\gamma_N = \gamma_N^{0} \exp(- \left| {\bf k} - {\bf K}^{(\prime)} \right| / k_r)$, 
introducing a damping factor which makes the term small
away from the Brillouin-zone corners, where this form
for the strain Hamiltonian becomes inaccurate, by setting  
$k_r = \gamma_1 / \hbar \upsilon_F = 0.0573 \AA^{-1}$.

Because there is some confusion in the literature on the signs of the remote hopping processes,
we have also considered other sign choices for $\gamma_2$ and $\gamma_3$.  
As shown in Fig. 1(b) direct hopping $\gamma_2$ between the low energy sites $A$, $B^{\prime \prime \prime}$
gives rise to three Fermi points at the vertices of a triangle centered on the 
Brillouin-zone corner.
The trigonal warping ($\gamma_3$) process which connect the $A$, $B^{\prime}$ and 
$A^{\prime} $, $B^{\prime \prime}$ sites is also responsible for a trigonal distortion 
that leads to four Fermi points near K, as in bilayer graphene.
Each one of the three Fermi points contribute to a phase winding of $2 \pi$
for a total $6 \pi$ phase winding along paths that encircle all three points,
as expected in ABC trilayer graphene \cite{hongki1}.
(We use the term Fermi point to refer to a band crossing that is tied to the Fermi 
level of a neutral ABC trilayer.  The band crossing points are exactly at the 
Fermi level because we have neglected particle-hole symmetry breaking terms 
in the band structure model.)  
Both $\gamma_2$ and $\gamma_3$ terms break circular symmetry near the Dirac point by splitting a 
single Fermi point with cubic band dispersion into multiple Fermi points with linear dispersion. 
The orientations of the triangular distortion due to $\gamma_2$ and  $\gamma_3$
are opposite when both hopping parameters have the same sign.
First principles band structure calculations suggest that $\gamma_2$ dominates 
over $\gamma_3$ and determines the shape of the bands near the Dirac point.
(Note that $\gamma_2$ has a much greater influence on the two low-energy states than 
$\gamma_3$ for a given numerical value because it couples them directly, whereas 
$\gamma_3$ acts virtually via high-energy states.)   
When both terms are present simultaneously and have the same sign
the band structure can have a a hybrid shape; 
for some parameters values the bands consist of two 
intertwined triangles with opposite orientations that can exhibit up to nine Dirac cones.
The additional parameter, $\gamma_N$ couples $A$, $B^{\prime}$ and $A^{\prime}$, 
$B^{\prime \prime}$, like the $\gamma_3$ term, but without an accompanying factor $f({\bf k})$.
The $\gamma_N$ term qualitatively describes a  band deformation that lower the crystal rotational symmetry,
and is similar to the model used to mimic a small layer-sliding structural deformation in bilayer graphene \cite{kruczynski}.
This term is also useful to seed lowered rotational symmetry 
gapless states in our Hartree-Fock calculations.

Electron-electron interaction effects are treated in an unrestricted Hartree-Fock approximation \cite{dirachf,jeilbilayer}
which allows symmetries to be broken:  
\begin{eqnarray}
\label{hfgen}
V_{HF} &=& \sum_{{\bf k} \lambda \lambda^{\prime}} U_H^{\lambda \lambda^{\prime}}
\left[ \sum_{{\bf k}^{\prime}}
\left<  c^{\dag}_{{\bf k}^{\prime} \lambda^{\prime}} c_{{\bf k}^{\prime} \lambda^{\prime}} \right>  \right]
c^{\dag}_{{\bf k} \lambda} c_{{\bf k} \lambda}   \nonumber  \\  %
&-& \sum_{{\bf k}^{\prime}\lambda \lambda^{\prime}} U_{X}^{\lambda \lambda'}
\left({\bf k}^{\prime} - {\bf k} \right)
\left<  c^{\dag}_{{\bf k}^{\prime} \lambda^{\prime}} c_{{\bf k}^{\prime} \lambda} \right>
c^{\dag}_{{\bf k} \lambda} c_{{\bf k} \lambda^{\prime}} 
\end{eqnarray}
where $c^{\dag}_{{\bf k} \lambda}$, $c_{{\bf k} \lambda}$ are Bloch state creation and 
annihilation operators, and $\lambda = (l,\sigma)$ lumps lattice and spin indices. 
The Hartree and Exchange Coulomb integrals in Eq.(~\ref{hfgen}), 
\begin{eqnarray}
U_H^{l l^{\prime}} &=&  \frac{1}{A} \sum_{\bf G} e^{ i {\bf G} \cdot \left({\bf s}_{l} - {\bf s}_{l^{\prime}} \right) } 
\left| \widetilde{f}\left( \left| {\bf G}   \right|\right) \right|^2 \,\, \widetilde{V}^{l l^{\prime}} \left( \left|  {\bf G}   \right|  \right)  \label{mom1}
\\
U_X^{l \, l^{\prime}} \left( {\bf q} \right)    
&=& \frac{1}{A} \sum_{\bf G} \; e^{  i {\bf G} \cdot \left({\bf s}_{l} - {\bf s}_{l^{\prime}} \right) } 
\left| \widetilde{f} \left( \left| {\bf q}  -  {\bf G}  \right|\right) \right|^2 \,\, \widetilde{V}^{l l^{\prime}}
\left( \left|  {\bf q} - {\bf G}  \right|  \right), \quad \quad   \label{mom2}
\end{eqnarray}
involve sums over reciprocal lattice vectors ${\bf G}$.
In these equations ${\bf s}_l$ is the (2D projection of the) position of the sublattice in the unit cell.
We used an isotropic atomic orbital form factor 
$\widetilde{f}({\bf q}) = \int d{\bf r} \, e^{- {\bf q} {\bf r}}  \left| \phi \left( {\bf r} \right) \right|^2  
= (1 -  \left(r_o q\right)^2 ) / (  (1 + \left(r_o q \right)^2 )^4  )$
with an artificially large atomic radius $r_o = 3a_o/\sqrt{30}$ to account for 
$sp_2$ orbital polarization.\cite{dirachf}
Here $a_o = a/\left(2 \sqrt{3}\right)$ is the covalent bond radius of carbon.
The two-dimensional Coulomb interactions in  Eqs. (\ref{mom1}-\ref{mom2}) are defined by
$\widetilde{V}^{l l^{\prime}} \left( {\bf q} \right) =  2 \pi e^2 / \left(  \left| \bf q \right|   \epsilon_r \right)  $
when the sublattice indices $l$ and $l^{\prime}$ refer to the atoms in the same layer, and 
$\left( 2 \pi e^2 /\left( { \left| \bf q \right|}   \epsilon_r \right)  \right)  \exp{ \left[  - \left| {\bf q} \right|  d \right] } $
when they refer to atoms in layers separated by a distance $d$.

We used an effective dielectric constant $\varepsilon_r = 4$ in our 
calculations, partly to account for dielectric screening by surrounding material and 
partly to account for the well known tendency of Hartree-Fock 
approximation, which neglects screening, to overestimate exchange interaction effects.\cite{exchangereduc}
The present implementation of the lattice model Hartree-Fock mean-field theory 
follows closely the method described in Refs. [\onlinecite{dirachf,jeilbilayer}] for 
single and bilayer graphene which also used a momentum space representation
of the Coulomb interaction.  
One difference in the present implementation is that we sample the full 
Brillouin zone without taking advantage of the symmetry of the crystal in order
to allow for the possibility of broken rotational symmetry nematic phases.
Because of the greater importance of states near the Dirac point we have 
sampled momentum space non-uniformly; for k-points closer than
$\sim 0.5/a$ to the Dirac point (where $a = 2.46 \AA$ is the triangular 
lattice constant of graphene) 
we have used a sampling density corresponding to 2304 $\times$ 2304
points in the entire Brillouin zone.
Outside this region we used a matched but coarser sampling with density
corresponding to 18 $\times$ 18 points in the 
Brillouin zone.
For a given sampling density, the Hartree-Fock equations are 
solved iteratively and converged to $\sim 10^{-11}$ eV per carbon atom in total energy.

\begin{figure}[t]
\begin{center}
\includegraphics[width=8.5cm,angle=0]{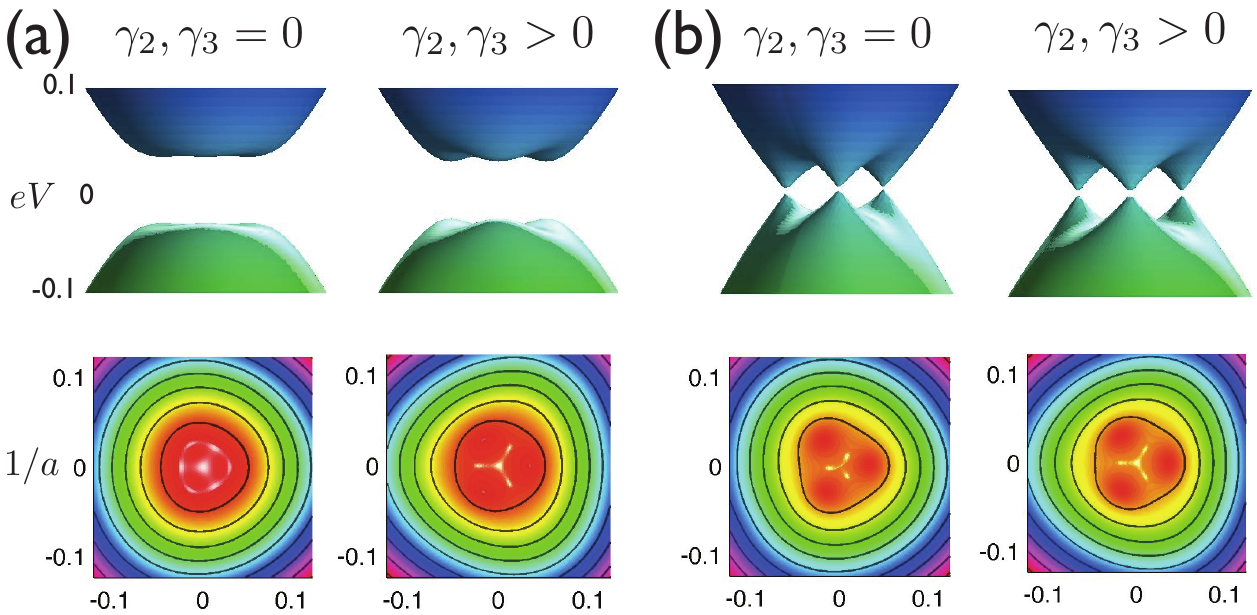} 
\caption{
Self-consistent Hartree-Fock calculations iterated from seeds
for the gapped and gapless nematic  
states in ABC trilayer graphene. 
The gapped solution (a) has been obtained starting from the layer
antiferromagnetic initial condition, while the gapless self-consistent solution (b) has been
obtained by seeding with a nematic $\gamma_N$ term. 
Note that the gapless solutions
restore the rotational symmetry of the crystal lattice
that was broken by the $\gamma_N$ term, indicating that nematic order is not stable at the 
mean-field level in the trilayer case.
When the remote hopping parameters $\gamma_2 = 0.01$ eV and $\gamma_3 = 0.3$ eV
are accounted for they induce a triangular distortion of the bands near the Dirac point
and determine the angles at which the band crossings occur.  These processes do not 
have a large influence on the gapped ground state.  
}
\label{energy}
\end{center}
\end{figure}

\begin{table}[ht] 
\centering 
\begin{ruledtabular}
\begin{tabular}{| c |  c c c c | c c c c |} 
\multicolumn{1}{|c|}{ AF}        &
\multicolumn{4}{c}{ $\left( \;\; \lambda_z \;\; \;\;\; \tau_z \;\; \;\;\;  \sigma_z  \;\; \right) $}        &
\multicolumn{1}{|c} { $\sigma^{K, \uparrow}_{xy}$ }       &      
\multicolumn{1}{c} { $\sigma^{K', \uparrow}_{xy}$ }     &       
\multicolumn{1}{c} { $\sigma^{K, \downarrow}_{xy}$ }       &      
\multicolumn{1}{c|} { $\sigma^{K', \downarrow}_{xy}$ }      \\   \hline 
AH &    $ (T \,\, K  \uparrow)$   &   $(B\,\,  K'  \uparrow)$    &  $(T \,\, K  \downarrow)$         &   
$(B \,\, K'  \downarrow)$   &  3    & 3 &  3  &  3    \\
SH &    $ (T \,\, K  \uparrow)$   &   $(B \,\, K'  \uparrow)$    &   $(T \,\, K'  \downarrow)$    & 
$(B\,\,  K  \downarrow)$     &  3   & 3  & -3  & -3    \\
LAF  &    $ (T \,\, K  \uparrow)$    &   $(T\,\,  K'  \uparrow)$   &   $(B \,\, K  \downarrow)$     &  
$(B\,\,  K'  \downarrow)$     &  3   & -3    & -3  &  3   \\
\end{tabular}

\smallskip

\begin{tabular}{|c| c c c c c c | } 
 & $\Delta n_l^{LAF}$ & $\Delta n_l^{AH}$ & $\Delta_{ gap}^{LAF}$    & $\Delta_{ gap}^{AH}$  & 
 $\Delta E_{cond}^{LAF}$  & $\Delta E_{cond}^{AH}$ \\ [0.5ex] 
\hline 
$ \gamma_{2}, \gamma_{3}      = 0$    &  1.22   &  1.21   &   65.1  &  64.9  & - 3.599 & - 3.554 
    \\ 
$ \gamma_{2}, \gamma_{3} > 0 $        &  1.09  &   1.08  &   56.0  &  55.6  & - 1.716  & - 1.680  \\
$ \gamma_{2} < 0, \gamma_{3} >0 $   &  0.12  &   0.10   &   12.1  &  11.7  & 0.00039 & 0.00046  \\ 
\end{tabular} 

\smallskip

\begin{tabular}{|c| c c c c | } 

& $\Delta E_{tot}$  & $\Delta E_{X}$ & $\Delta E_{X}^{KK}$  & $\Delta E_{X}^{K K^{\prime}}$ \\
\hline
$ \gamma_{2}, \gamma_{3}      = 0$     &   $ -4.43  $  & $-14.02 $   & $-2.62 $ & $-0.88 $     \\ 
$ \gamma_{2}, \gamma_{3} > 0 $         &   $ -3.58 $   & $ -13.81 $   & $ -2.77 $ & $ -0.68 $     \\ 
$ \gamma_{2} < 0, \gamma_{3} > 0 $   &   $ -0.0065 $   & $ -1.660 $   & $ -0.4154 $ & $ 0.0005 $   \\ 
\end{tabular} 
\end{ruledtabular}
\caption{ 
{\em Upper panel:}
Mean-field theory properties of the three balanced-charge-density gapped states.  
Each of these states has two of the four valley/spin 
flavors polarized towards the top layer and two toward the bottom layer.
Each polarized flavor contributes three 
quantized $e^2/h$ units to the Hall conductivity 
with a sign that depends on both valley and layer polarization;
the continuum model assignments can be retained in a lattice model because the 
momentum space Berry curvatures are strongly localized near Brillouin-zone corners.  
{\em Middle panel:}
The density transfer from one outer layer to the other $\Delta n_l$ for each valley-spin 
degree of freedom in units of $10^{11} cm^{-2}$.
(This density scale corresponds to $\sim 1.3 \cdot 10^{-5}$ electrons per carbon atom.)
The total amount of charge transferred per valley-spin is larger 
in the more stable LAF configuration than the AH or SH configurations.
 $\Delta_{gap}$ is the energy gap in meV.
 The condensation energies $\Delta E_{cond} = E_{gapped} - E_{gapless}$
 shown are differences between  the ordered gapped and gapless normal
 phases in units of $10^{-7}$ eV per carbon atom.
 The gapless normal state energies have been obtained from a self-consistent calculation
 starting from the band orbital seed.
 The anomalous Hall and spin Hall states have the same energy in 
mean-field theory.  
{\em Lower panel:}
Differences in total energy between LAF and AH/SH states, $\Delta E = E^{LAF} - E^{AH/SH}$, 
in units of $10^{-9}$ eV per carbon atom.  The exchange energy difference per spin/valley
is separated into an intravalley ($ \Delta E_{X}^{KK}$) and an intervalley ($ \Delta E_{X}^{KK'}$) contribution.   
Note that the total exchange energy difference satisfies  $\Delta E_X = 4 ( \Delta E_{X}^{KK} + \Delta E_{X}^{KK'} )$.
Intervalley exchange, normally neglected in continuum models,
makes a substantial contribution to the 
energy difference between LAF and anomalous Hall states.
}
\label{table:nonlin}
\end{table} 

\section{Gapped and gapless states} 
As in the AB bilayer case, the low energy valence band states of ABC graphene are 
given approximately by equal weight coherent sums of top and bottom layer 
wavefunctions with momentum-dependent phase differences.  
The gapped broken symmetry states spontaneously increase weight in one of the 
two layers, whereas\cite{nobelsymposium} the nematic states break the lattice rotational symmetry of the inter-layer 
phases.  In the following we present  the results of our $\pi$-band Coulomb-interaction 
Hartree-Fock study.  This mean-field-theory calculation we perform cannot be fully quantitative
because it accounts for screening in an ad-hoc way which might be quantitatively inaccurate,
and because it neglects higher-order correlation effects.  We believe though that our 
results provide some insight into the competition between different potential 
ordered states, and in particular into the way this competition is influenced by 
band structure features particular to ABC trilayer graphene.  We first discuss the gapped states,
which have spontaneous layer polarizations with spin or valley dependent signs, and then 
ungapped states with lowered rotational symmetry.    

In a continuum model, the energy of the gapped states is minimized when half of the
spin-valley components are polarized toward one layer and half to the other. \cite{jeilbilayer,fan}
We consider only states of this type, which are favored over other closely related states 
by electrostatic interactions.  
In a lattice model there is a clear distinction and an energy difference
between states with opposite layer polarizations
for different valleys, which have either an anomalous Hall (AH) effect or a spin Hall (SH)
effect, and states with opposite layer 
polarization for opposite spins (LAF), which form an antiferromagnetic state.  
The three types of ABC trilayer gapped states that have no overall layer polarization are
compared in Table I.  In our mean-field calculations anomalous Hall and spin Hall states 
have the same energy.

We define the condensation energy of the LAF and AH/SH gapped states 
as their energy relative to the ground state energy of the 
unbroken symmetry states.  
The unbroken symmetry state energy is determined by 
carrying out self-consistent mean-field calculations that are 
seeded by the non-interacting electron ground state.
We find that the condensation energies for the 
ordered states are $\sim 10^{-7}$ eV per carbon atom.
The condensation energy is approximately three times smaller than the 
product of the energy gap $\Delta_{gap}$ and the 
charge transferred between layers within individual spins and valleys $\Delta n_l$. 

The condensation energy scales are approximately five times 
larger than those obtained for bilayer graphene\cite{jeilbilayer}
with similar approximations, presumably because the crossing bands are 
even flatter in the trilayer case, increasing the role of interactions.     
The band gaps we calculate and present in Table I are roughly ten times larger than the 
the spontaneous gap values $\sim 6$ meV estimated from transport 
measurements in ABC trilayer graphene \cite{trilayers1}, 
and between 1.6 to 2 times larger than the gaps ($\sim 30$ meV) obtained for the bilayer graphene
using the same value of $\epsilon_r = 4$. [\onlinecite{jeilbilayer}]
This could suggest that screening effects are underestimated by this value of 
$\epsilon_r$, or that other interaction effects than are absent in 
mean-field-theory play an essential role.  

Experimentally the ratio of trilayer to bilayer gaps is $\sim 2.5$,
close to the ratio we obtain.   
This suggests that the choice $\epsilon_r=4$ quantitatively
overestimates exchange effects in both cases. 
The discrepancy between theory and experiment could, however, be 
due in part to the unfavorable influence of disorder in experimental
 samples, and also in part to inaccuracies in our band structure model.  
From the results in Table I we can observe, for example, 
that the gapped states are strongly suppressed
when $\gamma_2$ and $\gamma_3$ have opposite signs, separating 
the Fermi points of the three Dirac cones. 
On the other hand, when $\gamma_2$ and $\gamma_3$ have the same
sign, \cite{tightbindingtrilayer}
the overall effect is that of restoring the approximate circular symmetry of the bands,
enhancing the chances for a gapped phase.

Our calculations find that the nematic broken symmetry state is not stable in ABC trilayers.
When we iterate the Hartree-Fock equations starting from a nematic seed, 
lattice rotational symmetry is restored at convergence.  
In both $\gamma_2=\gamma_3=0$,
and the more realistic $\gamma_2,\gamma_3 \ne 0$ case, the 
same unbroken symmetry state with three band crossing points
is reached for self-consistent 
calculations starting from either nematic or band seeds.   
This result is different from the one obtained in the graphene bilayer case, in which the same type of calculation 
yields a stable gapless state which lowers the crystal's rotational symmetry giving rise to a 
nematic order. \cite{inprep}

The gapped solution of the Hartree-Fock equations  
lowers the total energy of the system by avoiding rapid in-plane $xy$ rotation of 
the sublattice pseudospin direction near the 
band crossing point.\cite{hongki}
The gapless nematic phase lowers the total energy of the system by
reducing the wavevector dependence of inter-site phase differences and 
introduces an anisotropic renormalization of the band velocity.
The competition between the two broken symmetry phases 
depends on how much energy can be gained by 
reshaping the quasiparticle bands in two different ways.  
Fig. 2 shows the band structures obtained from self-consistent  
calculations with gapped and nematic seeds.
\begin{figure}[t]
\begin{center}
\begin{tabular}{c}
\includegraphics[width=8.5cm,angle=0]{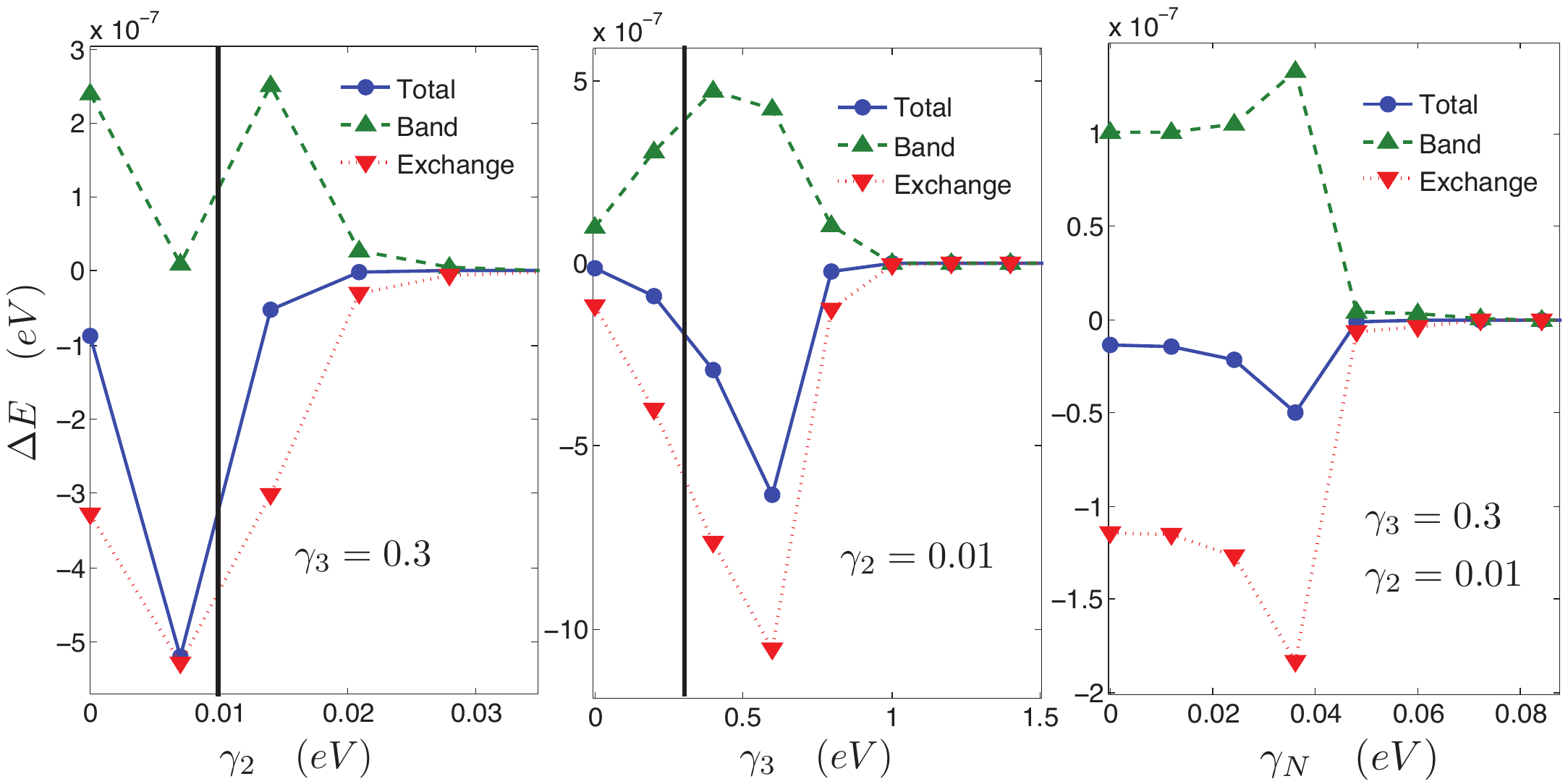}  \quad \quad \quad \quad
\end{tabular}
\caption{(Color online)
Energy difference between gapped and gapless states $\Delta E = E_{gapped} - E_{gapless}$
as a function of $\gamma_2$, $\gamma_3$ and $\gamma_N$.
For $\gamma_N=0$ the lattice symmetry of the gapless state is not lowered by 
interactions.  The vertical black solid lines indicate the hopping parameters 
$\gamma_2 = 0.01$ eV and $\gamma_3 = 0.3$ eV
that best approximate the band structure predicted by {\em ab initio} DFT calculations.
For strong remote hopping processes, the gap in the gapped state closes progressively and
the energy difference between gapped and ungapped states is reduced progressively.
}
\label{energy}
\end{center}
\end{figure}
The remote hopping terms introduce a 
triangular distortion in the shape of the bands near the Fermi energy, 
but do not greatly influence gapped state properties.  
These distortions will have important consequences for the electronic properties
of doped ABC trilayers.  
It is noteworthy that the gapless phase within the minimal model develops
a three Fermi point structure due to electron interactions alone, although
it does not lower rotational symmetry.  
The inclusion of the $\gamma_2, \gamma_3$ terms also plays a role in 
determining the orientation of the triangular deformation 
the bands undergo near the Dirac points in the gapless state.

Motivated by uncertainty in the values of the remote hopping process 
parameters, we have performed self-consistent calculations 
over a range of values of the $\gamma_2$, 
$\gamma_3$ and $\gamma_N$ parameters.
The dependence of the energy difference between the interaction driven 
gapped and gapless states on model parameters is summarized in 
Fig. 3.  We find that the gapped phase almost always has a lower total energy
than the gapless phase. However, as expected, when the
remote hopping processes are stronger,  
the difference in the total energy between the gapped and gapless phases
become smaller.  Fig. 3 shows that the occurrence of the gapped phase relies on the 
principal intra and interlayer processes whose strength is defined by 
$\gamma_0$ and $\gamma_1$, and by the flatness of the
crossing between conduction and valence bands that their dominance implies.

\section{Discussion}

We have used Hartree-Fock mean-field-theory calculation 
to demonstrate that electron interactions can lead to ordered 
phases in ABC trilayer graphene, provided that the strengths of 
remote hopping process in this two-dimensional crystal are close 
to current estimates.
In ABC trilayers, bands near the Dirac point are strongly influenced 
by the $\gamma_2$ parameter over energy scales of $\sim 20$ meV,
compared to the $\sim 1$ meV scale over which analogous processes
play a role in AB bilayers.  The physics of their interplay with interactions 
is therefore less likely to be distorted by disorder.
Remote hopping processes in ABC trilayers
can be important in fixing the shape of the energy bands near the Fermi level.  We have shown that the 
gapped broken symmetry phases are nevertheless 
preferred energetically over gapless states
for a wide range of remote hopping parameters. 
We find that our gapless solutions do not lower the crystal symmetry,
although they do generally lead to the formation of a triple Dirac point at the vertices of an equilateral
triangle.  The nematic phase that would break the triangular crystal symmetry
is not stable.
When remote hopping processes are included the location 
of the Dirac points is fixed by the $\gamma_2$ process.

There are three distinct gapped states which have very similar 
energies.  Among these the anomalous Hall and spin Hall (AH and SH) states 
have the same energy within mean-field theory, whereas the layer 
antiferromagnet state is distinct and is favored by inter-valley exchange because 
electrons with the same spin state have the same sense of layer polarization.   
We find that the difference in total energy
between LAF and  
AH/SH is two orders of magnitude smaller than the condensation 
energy of either state.  These states should therefore, in our view,
be considered as close cousins.  It seems likely that real samples are
likely to be found in configurations in which several phases are present 
separated by domain walls.  An external magnetic field which favors anomalous 
Hall states, at least at finite carrier densities, likely can be used to 
manipulate the domain structure.  

Using the Hartree-Fock approximation and reducing interaction strengths by a 
factor of $\epsilon_r = 4$, we find that ABC trilayer graphene has a substantial interaction 
driven gap of the order of $65$ meV.   The size of the gap is sensitive to the 
choice we have made for the $\epsilon_r$ parameter, which we have chosen to 
mimic exchange interaction renormalization parameters that are used in 
{\em ab initio} hybrid-density-functional calculations.   Band structure effects can reduce the size of the 
gap, substantially so when $\gamma_2$ is assigned a negative value. 
For favorable parameters the gaps we find are approximately twice as large as  
than those predicted values for bilayer graphene using corresponding approximations.
The theoretical gaps are therefore very much larger than 
initial estimates of a spontaneous band gap from ABC trilayer 
experiments which suggest a value
$\sim 6$ meV. \cite{trilayers1}  The discrepancy is certainly due in part to 
disorder and inhomogeneity which reduces the gaps of experimental systems 
below ideal values, but could also reflect a theoretical overestimate.  
We note in this connection that ABC trilayer graphene samples generally have poorer 
quality than bilayers.  This difference could be due to lower effectiveness of the 
current annealing procedure routinely applied to suspended graphene single or multilayer 
samples.  Future experimental work may establish a higher lower bound for the 
trilayer graphene gap.

\acknowledgements 

This work was supported by SWAN, by Welch Foundation grant TBF1473,
and by DOE Division of Materials Sciences and Engineering grant DE-FG03-02ER45958.

\end{document}